# NXgraph: An Efficient Graph Processing System on a Single Machine

Yuze Chi*, Guohao Dai*, Yu Wang*, Guangyu Sun[†], Guoliang Li* and Huazhong Yang*
*Tsinghua National Laboratory for Information Science and Technology, Tsinghua University
{chiyz12,dgh14}@mails.tsinghua.edu.cn,{yu-wang,liguoliang,yanghz}@tsinghua.edu.cn
[†]Center for Energy Efficient Computing and Applications, Peking University
gsun@pku.edu.cn

*Abstract*—Recent studies show that graph processing systems on a single machine can achieve competitive performance compared with cluster-based graph processing systems. In this paper, we present NXgraph, an efficient graph processing system on a single machine. With the abstraction of vertex intervals and edge sub-shards, we propose the Destination-Sorted Sub-Shard (DSSS) structure to store a graph. By dividing vertices and edges into intervals and sub-shards, NXgraph ensures graph data access locality and enables fine-grained scheduling. By sorting edges within each sub-shard according to their destination vertices, NXgraph reduces write conflicts among different threads and achieves a high degree of parallelism. Then, three updating strategies, i.e., Single-Phase Update (SPU), Double-Phase Update (DPU), and Mixed-Phase Update (MPU), are proposed in this paper. NXgraph can adaptively choose the fastest strategy for different graph problems according to the graph size and the available memory resources to fully utilize the memory space and reduce the amount of data transfer. All these three strategies exploit streamlined disk access pattern. Extensive experiments on three real-world graphs and five synthetic graphs show that NXgraph can outperform GraphChi, TurboGraph, VENUS, and GridGraph in various situations. Moreover, NXgraph, running on a single commodity PC, can finish an iteration of PageRank on the Twitter [1] graph with 1.5 billion edges in 2.05 seconds; while PowerGraph, a distributed graph processing system, needs 3.6s to finish the same task.

## I. INTRODUCTION

With explosion of data volume generated and collected from ubiquitous sensors, portable devices and the Internet, we are now moving into the "Big Data" era. There exist various modern Big Data applications relying on graph computing, including social networks, Internet of things, and neural networks. For example, Facebook has 1.44 billions monthly active users during the first quarter of 2015 [2]. Both user data and relationship among them are modeled by graphs for further exploration. To this end, it has become a very important problem to process, analyze, and understand these graphs.

In order to achieve scalable graph computing, researchers have proposed many distributed or single machine solutions [3]–[23]. Representative distributed systems include Power-Graph [18], Giraph [15], Pregel [16], GraphLab [17],GraphX [19], PEGASUS [20], and etc. Some of these systems are developed based on popular distributed computing frameworks, such as MapReduce [24] and Spark [25]. These existing distributed approaches have achieved impressive high performance. For example, it takes PowerGraph [18] only 3.6s to execute the PageRank [26] algorithm per iteration on the Twitter [1] graph which is composed of 1.5 billions edges.

Performance of a distributed solution highly relies on the underlying infrastructure, which is composed of multiple computing nodes, distributed storage systems, and communication network among them. Distributed systems suffer from load imbalance, communication overhead and poor robustness. The only advantage over single-machine systems besides the performance is that a distributed system can scale to nearly arbitrarily large graphs whereas a single-machine system is limited by its computational resources. In fact, recent studies show that single-machine systems are able to handle graphs with billions of edges. For example, Twitter [1] graph has 42 millions of vertices and 1.5 billions of edges, which requires only 12 gigabytes to store if each edge is represented by 8 bytes. For these graphs, graph processing on a single machine can achieve comparable performance as a distributed system. For instance, GraphChi [22] takes 790s to finish the PageRank algorithm on Twitter, while Spark [25] takes 487s. Previous research has demonstrated that single machine based graph computing systems, like GraphChi [22], VENUS [21], TurboGraph [23], MapGraph [27], and X-stream [28] achieve comparable performance but with higher energy efficiency.

Because of poor locality [29], graph processing on a single machine faces challenges caused by the random access pattern. GraphChi [22] presents a novel Parallel Sliding Windows (PSW) model on a single machine. With the PSW model, GraphChi achieves a streamlined disk access pattern and addresses the locality problem. GraphChi provides a basic programming model for graph computation on a single machine. Subsequent researches follow this model and improves the system performance by introducing optimization techniques.

In the GraphChi system, there is a limitation that **all** incoming and outgoing edges of vertices in an interval need to be loaded into memory before calculation. This results in unnecessary disk data transfer. TurboGraph [23] addresses this problem with the *pin-and-slide* model, which also helps exploiting locality. VENUS [21] proposes an extra data structure, *v-shards*, to enable streamlined disk access pattern and high degree of parallelism. With two different Vertex-centric Streamlining Processing (VSP) algorithms, VENUS can either reduce the amount of disk data transfer or exploit the locality of data access. GridGraph [30] uses a *2-Level Hierarchical Partitioning* scheme to reduce the amount of data transfer, enable streamlined disk access, and maintain locality.

In general, a single machine system should mainly focus on the following four optimizing rules:

1) Exploit the locality of graph data.

2) Utilize the parallelism of multi-thread CPU.

3) Reduce the amount of disk data transfer.

4) Streamline the disk I/O.

Previous work addresses some aspects of the above to some extent, but none of the previous work addresses all the four aspects of the problem thoroughly. In this paper, we design NXgraph, following all these rules to improve the overall system performance. The main contributions of NXgraph are summarized as follows.

- **Destination-Sorted Sub-Shard (DSSS) structure:** To exploit locality of graph computation, both source and destination vertices need to be restricted to a limited range of memory space. NXgraph proposes a Destination-Sorted Sub-Shard (DSSS) structure which divides vertices and edges in a graph into intervals and **sub-shards**, respectively. With the sub-shards, graph data access locality is ensured and fine-grained scheduling is enabled. Meanwhile, NXgraph sorts edges within each sub-shard according to their **destination** vertices, which reduces write conflicts among different threads. Thus, NXgraph can achieve a high degree of parallelism. Experimental results show that sorting the edges by destinations achieves up to 3.5x speedup compared with sorting by sources.

- **Adaptive updating strategies:** To reduce the amount of disk data transfer and ensure streamlined access to the disk, we propose NXgraph with three updating strategies for graph computation, Single-Phase Updating (SPU), Double-Phase Updating (DPU), and Mixed-Phase Updating (MPU). SPU applies to machines with large memory space and minimizes the amount of disk data transfer. DPU applies to machines with small memory space. MPU combines the advantages of both SPU and DPU. All these three strategies exploit streamlined disk access pattern. We quantitatively model the updating strategies and analyze how to select a proper one based on the graph size and available memory resources.

- **Extensive experiments:** We do extensive experiments to validate the performance of our NXgraph system using both large real-world graph benchmarks and large synthetic graph benchmarks. We validate our design decisions first, followed by detailed experiments on different environments and different computation tasks. We also compare the NXgraph system with other state-of-the-art systems. Extensive experiments show that NXgraph can outperform GraphChi, TurboGraph, VENUS, and GridGraph in various situations.

The rest of this paper is organized as follows. The abstraction of graph computation model is shown in Section II. Based on Section II, the detailed system design of NXgraph is introduced in Section III. Section IV presents extensive experimental results. Section V introduces some previous work on graph processing. Section VI concludes the paper.

## II. COMPUTATION MODEL

### A. Graph Presentation

A graph $G = (V, E)$ is composed of its vertices $V$ and edges $E$. A computation task on $G$ is to read and update

TABLE I: Notations of a graph

| Notation | Meaning |
|---|---|
| $G$ | the graph $G = (V, E)$ |
| $V$ | vertices in $G$ |
| $E$ | edges in $G$ |
| $n$ | number of vertices in $G$, $n = |V|$ |
| $m$ | number of edges in $G$, $m = |E|$ |
| $v_i$ | vertex $i$ |
| $e_{i.j}$ | edge from $v_i$ to $v_j$ |
| $I_i$ | interval $i$ |
| $S_i$ | shard $i$, containing all edges whose destinations are in $I_i$ |
| $SS_{i.j}$ | sub-shard $i.j$, contains all edges whose sources are in $I_i$ and destinations are in $I_j$ |
| $H_{i.j}$ | hub $i.j$, contains all destination vertices and their attributes in $SS_{i.j}$ |
| $P$ | number of intervals |
| $Q$ | number of intervals that reside in memory |
| $B_a$ | size of a vertex attribute |
| $B_v$ | size of a vertex id |
| $B_e$ | size of an edge |
| $B_M$ | size of available memory budget |
| $d$ | average in-degree of the destination vertices of the sub-shards |

$V$ and $E$. We assume the graph to be directed. Updates are propagated from source vertex to destination vertex through the edge. Undirected graph is supported by adding two opposite edges for each pair of vertices.

To store vertices and edges on disk, vertices are organized as *intervals* and edges are organized as *shards*. All vertex values are divided into $P$ disjoint parts and each part is called an interval. All edges are then partitioned into $P$ shards, where each shard is associated with an interval. An edge belongs to a shard if and only if its destination vertex belongs to the corresponding interval. Moreover, each shard is further divided into $P$ *sub-shards* according to their source vertices. Inside each sub-shard, edges are sorted by their destination vertices. This structure of graph presentation is the Destination-Sorted Sub-Shard (DSSS). This presentation methodology of graph forms a two dimensional matrix of sub-shards, as shown in Figure 1. Notations used in this paper is listed in Table I.

For the example graph in Figure 1, interval $I_3$ consists of vertices $v_4$ and $v_5$. Interval $I_2$ consists of vertices $v_2$ and $v_3$. Therefore, sub-shard $SS_{3.2}$ consists of edges $e_{5.2}$, $e_{4.3}$, and $e_{5.3}$. When performing update on sub-shard $SS_{3.2}$, edges in $SS_{3.2}$ and vertex attributes in $I_3$ will be read and used to calculate new attributes for $I_2$. Here, interval $I_3$ is called the *source interval* as all source vertices reside in it and interval $I_2$ is called the *destination interval*. By restricting data access to the sub-shard and the corresponding source and destination intervals, locality is ensured under the DSSS structure.

### B. Update Scheme

A graph computation algorithm is composed of three parts: input, traversal and output. The input and output progresses are relatively straight-forward and will be addressed in the last part of this subsection. The tasks of the traversal progress are threefold. First, use old attributes stored in the intervals and the adjacency information in the sub-shards to calculate updated attributes of the intervals. Second, determine if the computation should be terminated. Third, maintain consistency within each iteration.

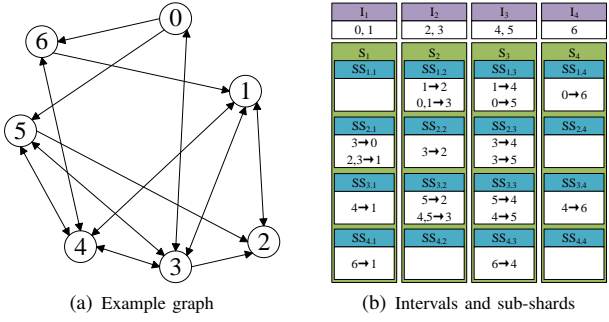

(a) Example graph  (b) Intervals and sub-shards

Fig. 1: Intervals and sub-shards in the graph

**Algorithm 1** NXgraph update model
---
**Input:** All intervals $I$ and sub-shards $SS$ of graph $G$, optional initialization data.
**Output:** Desired output results.
1: Initialize($I$)
2: **repeat**
3:   **if** all intervals is inactive **then**
4:     $finished \leftarrow$ **true**
5:   **else**
6:     $finished \leftarrow$ **false**
7:   **end if**
8:   $I_i \leftarrow$ inactive for all $I_i \in I$
9:   **for** each $SS_{i.j} \in SS$ **do**
10:     $I_j =$ Update($I_j, I_i, SS_{i.j}$)
11:   **end for**
12: **until** $finished =$ **true**
13: **return** Output($I$)

The first task of an iteration is to use the attributes in the source interval and the edges in the corresponding sub-shard to compute the new attributes of the destination interval. NXgraph performs the computation in unit of sub-shards, as shown in Algorithm 1.

The second task of an iteration is to determine whether to terminate the execution with *interval* activity status. If no vertex attribute in an active interval is updated in an iteration, that interval will be noted as inactive in the next iteration. Similarly, an interval will be activated if any vertex attribute in that interval is updated. When all the intervals enter the inactive state, the execution is determined to be terminated.

The last task of an iteration is to maintain consistency. NXgraph adopts a synchronous approach. Any attribute read from an interval must be the one written into it in the previous iteration. Write operation in the update scheme must not break this consistency.

The progress of input is to set initialization attributes and activity status for each interval according to the input. When the traversal is done, the intervals are traversed for the last one time to collect desired output.

For example, the input to an implementation of Breadth-First Search (BFS) is the root vertex. The initialization progress sets all vertex attributes to be infinity except that the root vertex is zero. Only the interval which contains the root vertex is active. The traversal progress update the destination vertex attribute with the minimum depth propagated from all its source vertices until no vertex can be updated. Finally, the output of the algorithm might be the maximum finite attribute of all intervals, which is the depth of the spanning tree given the specified root vertex. This example is shown as pseudo-codes in Algorithm 2, 3 and 4.

**Algorithm 2** BFS's Initialize($I$) function
---
**Input:** Intervals $I$, root vertex $v_{root}$.
**Output:** Initialized intervals $I$.
1: $I_i \leftarrow$ inactive for all $I_i \in I$
2: **for** each $v$ in $I$ **do**
3:   **if** $v$ is $v_{root}$ **then**
4:     $v.depth \leftarrow 0$
5:     $I_{root} \leftarrow$ active where $v_{root} \in I_{root}$
6:   **else**
7:     $v.depth \leftarrow \infty$
8:   **end if**
9: **end for**

**Algorithm 3** BFS's Update($I_j, I_i, SS_{i.j}$) function
---
**Input:** Destination interval $I_j$, source interval $I_i$ and sub-shards $SS_{i.j}$.
**Output:** Updated destination interval $I_j$.
1: **for** each $e \in SS_{i.j}$ **do**
2:   **if** $e.dst.depth > e.src.depth + 1$ **then**
3:     $e.dst.depth \leftarrow e.src.depth + 1$
4:     $I_j \leftarrow$ active if $I_j$ is inactive
5:   **end if**
6: **end for**
7: **return** $I_j$

**Algorithm 4** BFS's Output($I$) function
---
**Input:** Intervals $I$.
**Output:** Desired output result: maximum depth.
1: **return** $\max(v.depth)$ where $v \in I$

## III. SYSTEM DESIGN

### A. Preprocessing

As described in Section II-A, NXgraph uses intervals, sub-shards and an external initialization file as input. Therefore, NXgraph requires explicit preprocessing progress to generate the intervals and sub-shards used in updating.

**Degreeing.** The preprocessing is divided into two independent steps, *degreeing* and *sharding*. The degreeing step is only dependent on graph property. It maps the vertex *index* to continuous *id* and calculate the vertex degree. Here, index represents the number given in the raw input of a graph, may or may not have specific meaning and is possibly sparse. Id represents the unique identifier used to denote a vertex in NXgraph. Unlike indices, ids must be continuous. That is, a graph with $n$ vertices will be given indices from 1 to $n$. This procedure is used to provide constant-time access given a vertex id and eliminate non-existing vertices. The *degreer* generates a mapping file and a reverse-mapping file, which are used to obtain vertex id from index and obtain vertex index from id, respectively. It also generates a *pre-shard*, which is used as the input to the sharding step, where the vertex indices are already substituted to ids.

**Sharding.** After degreeing, the *sharder* will divide the pre-shard into $P^2$ sub-shards and allocate storage space for $P$ intervals. NXgraph does not have stringent limitations on the partitioning strategy except that at least one interval must be able to be stored in memory. In practice, we use much smaller interval size to enable adaptation to different memory sizes. We use an intuitive partitioning strategy which is to divide all vertices into equal-sized intervals. NXgraph adopts a fine-grained parallel model so that the possible imbalance among sub-shards will barely hurt system performance. Sub-shards are then generated according to the interval partitioning.

Because we use a continuous id as the identifier of a vertex, an interval can be stored as only attributes of vertices and an offset of the first vertex in the interval. This strategy not only reduces the access time to a vertex given its vertex id to constant time, but also decreases the space requirement to store an interval. As for sub-shards, unlike GridGraph [30], NXgraph sorts all edges by their destination vertex id so that edges with the same destination vertex is stored continuously. This enables efficient compressed sparse format of edge storage, which is beneficial to reduce the amount of disk I/O and enable fine-grained parallel computation. Furthermore, we also sort all edges with the same destination vertex in a sub-shard by their source vertex id, so that continuous memory segment will be accessed when performing update. This utilizes the hierarchical memory structure of CPU and helps maintain a high hit rate on CPU cache.

*B. NXgraph Update Strategy*

NXgraph proposes three novel update models, Single-Phase Update (SPU), Double-Phase Update (DPU), and Mixed-Phase Update (MPU). SPU is the fastest, but it requires that the memory is at least two times larger than all the intervals. Graphs with too many vertices will not be able to be processed with SPU. DPU is about two or three times slower, but it can adapt to small memories. DPU enables very large graph processing. MPU is a combination of the above two update strategies, which is a trade-off between memory requirement and system performance.

*1) Single-Phase Update:* **Description.** SPU is relatively straight-forward. SPU mode stores *two* copies of each interval in memory, one with the attributes from the previous iteration and the other with the attributes written into the interval in the current iteration. At the end of each iteration, the two copies will be exchanged so that the overhead of switching iteration is minimized. This procedure is called *ping-pong* operation of intervals. Edges stored in sub-shards are streamlined from either memory or disk, depending on whether memory space is sufficient. Before initialization, the SPU engine will actively allocate spaces for ping-pong intervals. If there are still memory budget left, sub-shards will also be actively loaded from disk to memory. As can be seen in Algorithm 1, *Update* function is performed in unit of sub-shards. In an iteration, each sub-shard $SS_{i,j}$ will invoke an *Update* function to update interval $I_j$ with previous attributes in interval $I_i$.

**Execution.** Since all intervals are present in memory, only sub-shards will need to be streamlined from disk. This guarantees sequential access to disk, which provides higher bandwidth than random access. Execution order among different sub-shards is not relevant in SPU mode. In practice, to avoid conflict upon destination intervals and maximize overlap between sub-shards, SPU will traverse the sub-shards by rows. With this traversal order, worker threads for the next sub-shard can be issued before all threads for the current sub-shard are finished. The pseudo-code for SPU mode is listed in Algorithm 5. These procedure corresponds to line 9 to 11 in Algorithm 1. Figure 2 illustrates the SPU schedule order with the example graph of Figure 1.

**Algorithm 5** An iteration with single-phase update
**Input:** Intervals $I$, sub-shards $SS$.
**Output:** Updated intervals $I$.
1: InitializeIteration($I_j$) for $j$ in $1 \to P$
2: **for** $i$ in $1 \to P$ **do**
3:   **for** $j$ in $1 \to P$ **do**
4:     $I_j = $ UpdateInMemory($I_j, I_i, SS_{i,j}$)
5:   **end for**
6: **end for**

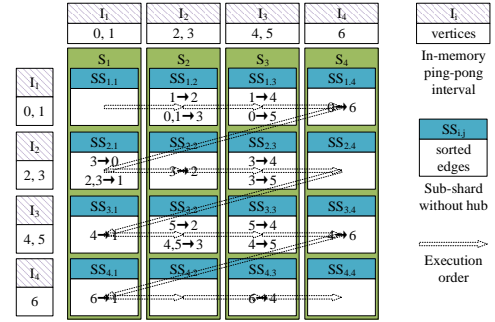

Fig. 2: Single-phase update order.

**Example.** In the example of Figure 2, SPU iterates over sub-shards $SS_{1.1}$ to $SS_{1.4}$ first. Computation on each of the sub-shards does not conflict with others. SPU can overlap the four sub-shards to fully exploit multi-threading. This corresponds to line 3 to line 5 in Algorithm 5. When computation on the first row of sub-shards is finished, SPU moves to the second row. This corresponds to the for loop in line 2 of Algorithm 5.

**I/O.** Consider a graph $G$ with $n$ vertices and $m$ edges, where $B_a$ bytes are used to store the attribute of a vertex and $B_e$ bytes are used to store an edge. Assume available memory budget is $B_M$ bytes where $2n \cdot B_a < B_M < 2n \cdot B_a + m \cdot B_e$. After initial load from disk, the amount of disk read per iteration will be at most:

$$B_{read} = m \cdot B_e + 2n \cdot B_a - B_M.$$

$B_{read} = 0$ if $B_M > 2n \cdot B_a + m \cdot B_e$. Since each edge will only be accessed once but each vertex will be accessed much more times in each iteration, it is more efficient to store intervals in memory than sub-shards. This memory utilization strategy minimizes the amount of disk I/O.

**Summary.** SPU strategy minimizes the amount of data transfer, optimizes both disk access pattern and memory access pattern, maximizes locality and provides high-degree and fine-grained parallelism. Therefore, its performance is maximized.

However, since at least two copies of all intervals need to be present in memory, SPU will be invalid for graphs with too many vertices. This issue will be solved by DPU and MPU mode, as will be addressed in the following paragraphs.

*2) Double-Phase Update:* **Description.** Unlike SPU, DPU mode is completely disk-based. Intervals will only be loaded from disk when accessed and sub-shards are streamlined from disk as in SPU mode. Since intervals can be partitioned to be much smaller than the total size of vertices, DPU can always handle very large graphs. DPU allocates a *hub* for each shard as an intermediate for vertex attributes. DPU shard hubs solve the consistency problem by store the update vertex attributes in them. A DPU shard hub consists of all destination vertex ids and their corresponding attributes. The attributes stored in a hub are incremental values contributed from its corresponding source interval, which can be accumulated later.

**Execution.** DPU reduces the amount of disk I/O by properly scheduling the updating order of the sub-shards. As indicated in its name, Double-Phase Update consists of two phases. The first is *ToHub* phase. In this phase, attributes in the previous iteration are read from the intervals and the calculated incremental attributes are written to the hubs. To avoid unnecessary disk reads, DPU iterates over the sub-shards by row in ToHub phase, loading each interval from disk only once per iteration, as shown in Figure 3. The second phase is *FromHub* phase. In this phase, attributes written into the hub are accumulated and written into the intervals. To minimize disk writes, DPU iterates over the sub-shards by column in FromHub phase, writing each interval to disk only once per iteration, as shown in Figure 4. The pseudo-code for DPU is in Algorithm 6.

---
**Algorithm 6** An iteration with double-phase update
---
**Input:** Intervals $I$, sub-shards $SS$.
**Output:** Updated intervals $I$.
1: **for** $i$ in $1 \to P$ **do**
2:     LoadFromDisk($I_i$)
3:     **for** $j$ in $1 \to P$ **do**
4:         $H_{i.j}$ = UpdateToHub($I_i, SS_{i.j}$)
5:     **end for**
6:     ReleaseFromMemory($I_i$)
7: **end for**
8: **for** $j$ in $1 \to P$ **do**
9:     InitializeIteration($I_j$)
10:    **for** $i$ in $1 \to P$ **do**
11:        $I_j$ = UpdateFromHub($I_j, H_{i.j}$)
12:    **end for**
13:    SaveToDisk($I_j$)
14: **end for**
---

**Example.** In the ToHub phase of DPU, as shown in Figure 3, DPU iterates over sub-shards $SS_{1.1}$ to $SS_{1.4}$ first. Interval $I_1$ is loaded into memory before the computation as shown in line 2 in Algorithm 6. Hubs $H_{1.1}$ to $H_{1.4}$ are written to the disk during the execution of the $UpdateToHub$ function in line 4. DPU can overlap the four sub-shards to fully exploit multi-threading since their write destinations, i.e., their hubs, do not overlap. When the computation on the first row is finished, DPU releases interval $I_1$ and loads the next interval $I_2$, as shown in line 6 and 2.

In the FromHub phase of DPU, as shown in Figure 4,

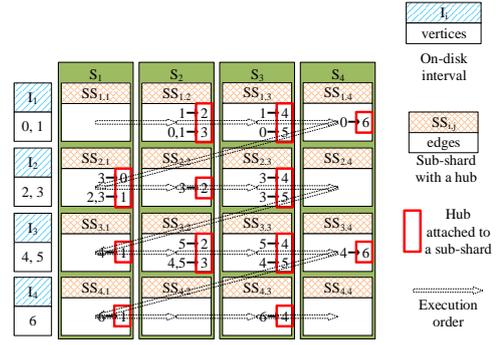

Fig. 3: ToHub phase execution order in DPU schedule.

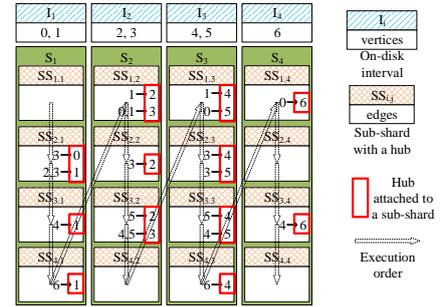

Fig. 4: FromHub phase execution order in DPU schedule.

DPU iterates over sub-shards $SS_{1.1}$ to $SS_{4.1}$ first. Interval $I_1$ is initialized in memory before the computation as shown in line 9. Hubs $H_{1.1}$ to $H_{4.1}$ are read from the disk during the execution of the $UpdateFromHub$ function in line 11 of Algorithm 6. When the computation on the first column is finished, DPU writes interval $I_1$ back to the disk and initializes the next interval $I_2$, as shown in line 13 and 9.

**I/O.** Consider a graph $G$ with $n$ vertices and $m$ edges, where $B_v$ bytes are used to store a vertex id, $B_a$ bytes are used to store the attribute of a vertex and $B_e$ bytes are used to store an edge. Assume $G$ is divided into $P$ equal-sized intervals and therefore $P^2$ sub-shards. In DPU mode, the amount of disk read and write per iteration will be at most:

$$B_{read} = m \cdot B_e + n \cdot B_a + m \cdot (B_a + B_v)/d$$
$$= m \cdot (B_e + \frac{B_a + B_v}{d}) + n \cdot B_a$$
$$B_{write} = n \cdot B_a + m \cdot (B_a + B_v)/d$$
$$= m \cdot \frac{B_a + B_v}{d} + n \cdot B_a.$$

In the above equations, $d$ denotes the average in-degree of the destination vertices of the sub-shards. For real-world graphs like Yahoo-web [31], typical value of $d$ is about 10 to 20, depending on the partitioning. Observe that $B_{read}$ and $B_{write}$ are not dependent on $P$ or memory budget, DPU mode can scale to very large graphs or very small memory budget without significant performance loss.

**Summary.** DPU can handle very large graphs with the price of high disk access compared to SPU. However, DPU is not the optimized trade-off point. As shown in Section IV-B3,

DPU is about two to three times slower than SPU. MPU mode will provide a better trade-off, as addressed in the next part of this section.

*3) Mixed-Phase Update:* **Description.** As analyzed in Section III-B1, it is more efficient to store vertices in memory than edges. Therefore, MPU loads as many intervals as possible in memory. For these in-memory intervals, MPU adopts the SPU-like strategy to perform update. As for those that cannot be loaded into memory at the same time, MPU uses the DPU-like strategy to perform update. Suppose $Q$ out of $P$ intervals reside in the memory. Only $(P-Q)^2$ out of $P^2$ sub-shards should use DPU-like update strategy whereas the rest of sub-shards can use SPU-like strategy. This is done as follows.

**Execution.** First, the MPU engine iterates over the $Q^2$ sub-shards using the same order of SPU. Then, the MPU loads each on-disk interval $I_i$ as the source interval and iterates over its corresponding sub-shards, $SS_{i,j}$. When the destination interval resides in memory, SPU-like update can still be performed since both source and destination intervals are loaded into the memory. When the destination interval resides on disk, the first phase, i.e., the ToHub phase is applied on the sub-shard. After that, the MPU engine will then iterate over sub-shards $SS_{i,j}$ where $0 < i \leq P, Q < j \leq P$ by column. When the source interval resides in memory, MPU still adopts the SPU-like strategy. When the source intervals resides on disk, MPU will perform the second phase of DPU, i.e., the FromHub phase on each sub-shard, updating the destination intervals with attributes stored in the shard hubs. The pseudo-code for MPU is Algorithm 7.

---

**Algorithm 7** An iteration with mixed-phase update

**Input:** Intervals $I$, sub-shards $SS$.
**Output:** Updated intervals $I$.
1: InitializeIteration($I_j$) **for** $j$ in $1 \to Q$
2: **for** $i$ in $1 \to Q$ **do**
3:   **for** $j$ in $1 \to Q$ **do**
4:     $I_j$ = UpdateInMemory($I_j, I_i, SS_{i,j}$)
5:   **end for**
6: **end for**
7: **for** $i$ in $Q+1 \to P$ **do**
8:   LoadFromDisk($I_i$)
9:   **for** $j$ in $1 \to Q$ **do**
10:     $I_j$ = UpdateInMemory($I_j, I_i, SS_{i,j}$)
11:   **end for**
12:   **for** $j$ in $Q+1 \to P$ **do**
13:     $H_{i,j}$ = UpdateToHub($I_i, SS_{i,j}$)
14:   **end for**
15:   ReleaseFromMemory($I_i$)
16: **end for**
17: **for** $j$ in $Q+1 \to P$ **do**
18:   InitializeIteration($I_j$)
19:   **for** $i$ in $1 \to Q$ **do**
20:     $I_j$ = UpdateInMemory($I_j, I_i, SS_{i,j}$)
21:   **end for**
22:   **for** $i$ in $Q+1 \to P$ **do**
23:     $I_j$ = UpdateFromHub($I_j, H_{i,j}$)
24:   **end for**
25:   SaveToDisk($I_j$)
26: **end for**

---

**Example.** Take the example in Figure 5. In this case, $P =$

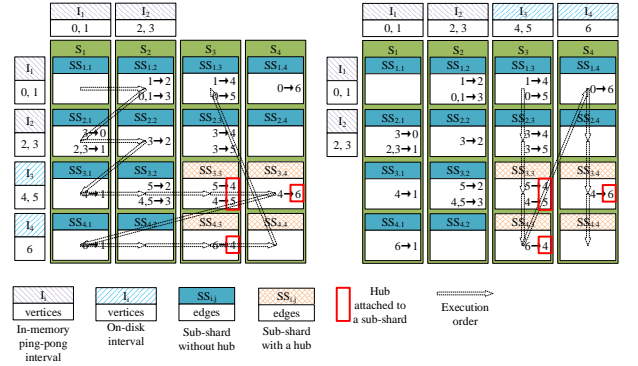

Fig. 5: Mixed-phase update order.

$4, Q = 2$, Interval $I_1$ and $I_2$ reside in memory. $I_3$ and $I_4$ are loaded into memory when accessed. Among the $P^2 = 16$ sub-shards, only $(P-Q)^2 = 4$ of them are attached a hub. MPU first updates sub-shards $SS_{1.1}$, $SS_{1.2}$, $SS_{2.1}$, and $SS_{1.2}$ with SPU, as shown in line 2 to 6 in Algorithm 7. Then MPU loads interval $I_1$ into memory and iterates over $SS_{3.1}$ to $SS_{3.2}$ as shown in line 8 to 11, performing SPU. On sub-shards $SS_{3.3}$ to $SS_{3.4}$, MPU performs the ToHub phase of DPU on them, as shown in line 12 to 14. After that, MPU iterates over the next row, as shown in the for loop of line 7. For column 3 of the sub-shards, as shown in the right sub-figure of Figure 5, MPU performs SPU on sub-shard $SS_{1.3}$ to $SS_{2.3}$ first. For sub-shard $SS_{3.3}$ to $SS_{4.3}$, MPU performs the FromHub phase of DPU. After that, MPU writes interval $I_3$ back to the disk (line 25) and moves to the next column, as shown in the for loop of line 17.

**I/O.** Take the assumptions in Section III-B2. Further assume that the sub-shards has the same size for the sake of simplicity. Assume available memory budget is $B_M$ bytes. Note that to perform SPU, these $Q$ intervals must be maintained as ping-ping intervals, which will take $2Q \cdot B_a$ bytes of memory space. In MPU, the amount of disk read and write per iteration will be at most:

$$B_{read} = m \cdot B_e + \frac{P-Q}{P} n \cdot B_a + (P-Q)^2 \cdot \frac{m}{P^2} \cdot (B_a + B_v)$$
$$= m \left[ B_e + \frac{(P-Q)^2}{P^2}(B_a + B_v) \right] + \frac{P-Q}{P} n \cdot B_a$$
$$B_{write} = \frac{P-Q}{P} n \cdot B_a + (P-Q)^2 \cdot \frac{m}{P^2} \cdot (B_a + B_v)$$
$$= \frac{(P-Q)^2}{P^2} m \cdot (B_a + B_v) + \frac{P-Q}{P} n \cdot B_a$$

where

$$Q \leq \frac{B_M}{\frac{2n}{P} B_a} = \frac{B_M}{2nB_a} P.$$

When $B_M > 2n \cdot B_a$, $Q = P$ and MPU will act exactly the same as SPU. When $Q = 0$, MPU will act exactly the same as DPU. When $B_M$ goes smaller, or equivalently, the scale of graph goes larger, I/O amount of MPU has the same upper bound as DPU.

**Summary.** Both SPU and DPU take advantages of sequential disk access, high memory access locality and high parallelism. As a combination of SPU and DPU, MPU has the above advantages as well. MPU also has the advantage of adaptation to different graph scales and low disk access amount. With properly partitioned intervals and shards, MPU can seamlessly adapt to different memory budgets without the need of re-doing preprocessing or significant loss on system performance. Therefore, NXgraph uses MPU by default.

*C. Comparison with TurboGraph-like Update*

For small memory sizes, we use MPU which combines SPU and DPU together. However, TurboGraph [23] and Grid-Graph [30] use another update strategy for small memory sizes. In this subsection, we will analyze the amount of data transfer of DPU and TurboGraph-like strategy, to illustrate why we don't combine SPU and TurboGraph-like updating strategy.

TurboGraph and GridGraph first load several source and destination intervals which can be fit into the limited memory. After updating all the intervals inside the memory, they replace some of the in-memory intervals with on-disk intervals. Turbo-Graph uses this strategy with active interval caching whereas GridGraph uses this strategy with the 2-level partitioning mechanism. We name this strategy after TurboGraph since it is an earlier work than GridGraph. We address only the I/O amount without considering interval caching mechanism, since the caching part is basically the same as MPU.

Take the assumptions in Section III-B3. Under the TurboGraph-like strategy, the amount of disk read and write per iteration will be:

$$B_{read} = m \cdot B_e + P^2 \cdot \frac{n}{P} \cdot B_a$$
$$= m \cdot B_e + nP \cdot B_a$$
$$B_{write} = P \cdot \frac{n}{P} \cdot B_a = n \cdot B_a$$

where

$$P \geq \frac{2nB_a}{B_M}.$$

As $P$ increases, the amount of I/O will increase linearly. This puts a limitation on partitioning. To get the best performance, the system must be partitioned into about $2nB_a/B_M$ intervals or the system will not run as fast as it could be. TurboGraph mentioned that it uses a page (interval) size of several megabytes, which will result in quite a number of pages, significantly damaging the performance. TurboGraph uses an active caching mechanism to mitigate this problem, but it still requires too much I/O to work efficiently on HDD. Grid-Graph uses a two-level partitioning mechanism to dynamically combine intervals and sub-shards together. This brings much flexibility since the equivalent $P$ can be significantly reduced, but it also loses the advantages of sorted sub-shards, namely, efficient compressed sparse format of edges storage and fine-grained parallelism. Besides, as the scale of graph increases, the combination mechanism will eventually stop working and $P$ will start to increase.

As a comparison, we use a real-world graph, Yahoo-web [31], as an example to show the different on the amount of disk I/O. For Yahoo-web, $n = 7.20 \times 10^8$ and $m = 6.63 \times 10^9$. Note that the vertex number here is less than the number of vertex indices because there are a large amount of vertices with no edge connected to it. When the memory is not sufficient for SPU, the best practice for MPU is to take $Q = \frac{B_M}{2nB_a}P$ and the amount of read and write per iteration will be:

$$B_{MPU} = m \cdot B_e + 2m \left(1 - \frac{B_M}{2nB_a}\right)^2 (B_a + B_v)/d$$
$$+ 2\left(1 - \frac{B_M}{2nB_a}\right) nB_a.$$

where $d$ is the average in-degree of the destination vertices in the $(P - Q)^2$ sub-shards with hubs. Experimental results indicate that this value is about 10 to 20 for Yahoo-web. We will use $d = 15$ for calculation. We assume a 4-byte vertex id and an 8-byte vertex attribute as it is for the PageRank algorithm. For edges, we assume an edge can be represented with about 4 bytes in average.

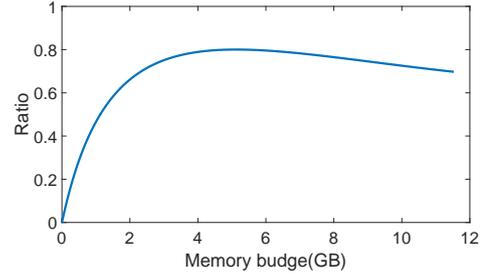

Fig. 6: Ratio of total I/O of MPU and TurboGraph-like.

As for TurboGraph-like strategy, the total amount of I/O is minimized when $Q = 0$ and $P = 2nB_a/B_M$. Therefore, the amount of read and write per iteration will be:

$$B_{TurboGraph\text{-}like} = m \cdot B_e + 2\frac{(nB_a)^2}{B_M} + nB_a.$$

The ratio of total I/O between MPU and TurboGraph-like strategy when memory budget varies from 0 to $2nB_a$ is shown in Figure 6. As the plot shows, MPU always out-performs TurboGraph-like strategy. When available memory budget grows even larger, SPU starts to be valid and will bring even more performance benefits.

We list all four update strategies addressed above in Table II. As can be seen in the table, SPU outperforms all other strategies on both read and write data amount. As for DPU, although it requires more disk writing, it has less amount of data transfer as analyzed above. The amount of data transfer of MPU is between SPU and DPU. When $B_M = 0$, MPU becomes DPU and when $B_M = 2nB_a$, MPU is the same as SPU.

*D. Fine-grained Parallelism in Each Sub-shard*

NXgraph adopts a fine-grained parallelism model within each Destination-Sorted Sub-Shard for both SPU and DPU as well as their combination, MPU. It exploits the power of multi-thread CPU as follows.

Since the edges in a sub-shard is ordered by destination vertex ids, the execution is issued in bunch of destination vertices. That is, within a sub-shard, each worker thread will take charge of several destination vertices and their associated edges. There is no write conflict between these threads since they are in charge of different destinations. No thread locks or atomic operations are required to maintain consistency except for the control signal. This enables high degree of parallelism when a sub-shard is large enough, i.e., to the scale of several thousands of edges. At the same time, access to the disk is still sequential as long as each thread reads and caches its own data. Besides, worker threads for different sub-shards can overlap with each other as long as their destination intervals are not the same. SPU takes advantage of this and exploits even more parallelism. In the case of DPU, everything is the same except that the destination of write operation becomes the hub instead of the interval for the ToHub phase. Because the destinations are sorted, the write operation to the hub is also sequential. For the FromHub phase, the execution order determines that threads cannot be overlapped among hubs, which will hurt performance to some extent. However, the high parallelism within a hub is still valid. GraphChi, TurboGraph or GridGraph cannot provide similar advantages.

TABLE II: Amount of read and write for different update strategies

|  | $B_{read}$ | $B_{write}$ |
|---|---|---|
| TurboGraph-like | $mB_e + 2(nB_a)^2/B_M + nB_a$ | $nB_a$ |
| SPU | $mB_e - (B_M - 2nB_a), B_M > 2nB_a$ | 0 |
| DPU | $mB_e + m(B_a + B_v)/d + nB_a$ | $m(B_a + B_v)/d + nB_a$ |
| MPU | $mB_e + m(1 - B_M/2nB_a)^2(B_a + B_v)/d + n(1 - B_M/2nB_a)B_a$ | $m(1 - B_M/2nB_a)^2(B_a + B_v)/d + n(1 - B_M/2nB_a)B_a$ |

Inside each worker threads, NXgraph takes advantage of CPU cache hierarchy by sorting edges with the same destinations according to their source vertex id. In this way, both spacial locality for the read operation and temporal locality for the write operation can be guaranteed. As a comparison, GridGraph maintains good locality by using a fine-grained lower-level partitioning strategy so that access to the main memory is limited to the lower-level of chunks, but that will make partitioning dependent on the size of CPU cache and its caching strategy, which is out of the designers' control. Besides, since the edges are not sorted, due to the multi-level hierarchy of CPU caches, the hit rate for lower-level caches will be lower than the sorted case. GraphChi and TurboGraph do not address the CPU cache locality issue. VENUS exploits locality in the dynamic view algorithm where v-shards are materialized, but suffers from the overhead of maintaining consistency among v-shard values.

## IV. EXPERIMENTS

In this section, we will first describe the evaluation setup of our experiments, followed by the demonstration on how system performance varies with different computational resources. We will then demonstrate system performance with different computational tasks. Further comparison with other state-of-the-art systems are listed in the next part of this section, followed by the analysis on how design factors should affect system performance.

Note that in the implementation of NXgraph update model, there can be two different mechanisms to solve the synchronization problem. One is to invoke a callback function at the end of each worker thread and send a proper signal.

TABLE III: Datasets used in the experiments

| Dataset | # Vertices | # Edges |
|---|---|---|
| Live-journal | 4.85 million | 69.0 million |
| Twitter | 41.7 million | 1.47 billion |
| Yahoo-web | 720 million | 6.64 billion |
| delaunay_n20 | 1.05 million | 6.29 million |
| delaunay_n21 | 2.10 million | 12.6 million |
| delaunay_n22 | 4.19 million | 25.2 million |
| delaunay_n23 | 8.39 million | 50.3 million |
| delaunay_n24 | 16.8 million | 101 million |

# vertices does not include isolated vertices.

NXgraph engine reads and controls the execution according to this synchronization signal. The other is to set a lock on each destination interval when writing, blocking the worker threads with write conflicts. These two implementations are both supported in current version of implementation and experimental results show that either one can always outperforms the other. Therefore, we report results for both of them in our experiments.

### A. Evaluation Setup

We run all our experiments on a personal computer equipped with a hexa-core Intel i7 CPU running at 3.3GHz, eight 8GB DDR4 memory and two 128GB SSD configured as RAID 0. This computer also has a 1TB HDD for operating systems. NXgraph and GraphChi are evaluated under Ubuntu 14.04 LTS 64bit version whereas TurboGraph is evaluated under Windows 10 64bit educational edition. Note that under Ubuntu, we are able to change memory size and CPU thread number at runtime by tweaking Linux kernel options but Windows does not offer similar interfaces. Therefore, for TurboGraph, we install only 2×8GB memories so that the memory resource is comparable with the other two systems. We cannot change the number of CPU threads for TurboGraph. We does not evaluate memory resource consumption with the memory budgets specified by the program because the operating system will use unoccupied memory space as disk cache, which will bring unpredicted performance impacts. We will always specify the maximum memory budget possible.

To get rid of the impact of random phenomenons, we run PageRank for 10 iterations unless otherwise specified. We limit the memory size to 16GB and enable all 12 available threads with hyper-threading unless otherwise specified. The properties of the datasets used in the experiments are listed in Table III. Note that vertices without any edge connected to them are excluded in the number of vertices.

### B. Design Decisions

*1) Exp 1 Sub-shard Ordering and Parallelism:* NXgraph sorts the edges in a sub-shard by their destination ids and adopts a fine-grained parallelism, as first proposed in VENUS

TABLE IV: Performance with different sub-shard model

| Model | Elapsed Time(s) | | |
|---|---|---|---|
| | Live-journal | Twitter | Yahoo-web |
| src-sorted, coarse-grained | 1.44 | 72.06 | 696.14 |
| dst-sorted, fine-grained | 1.00 | 20.50 | 519.31 |

Task: 10 iterations of PageRank

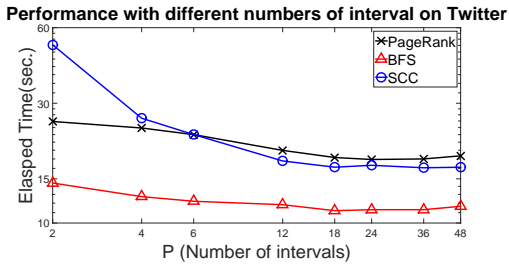

Fig. 7: Performance with different partitioning.

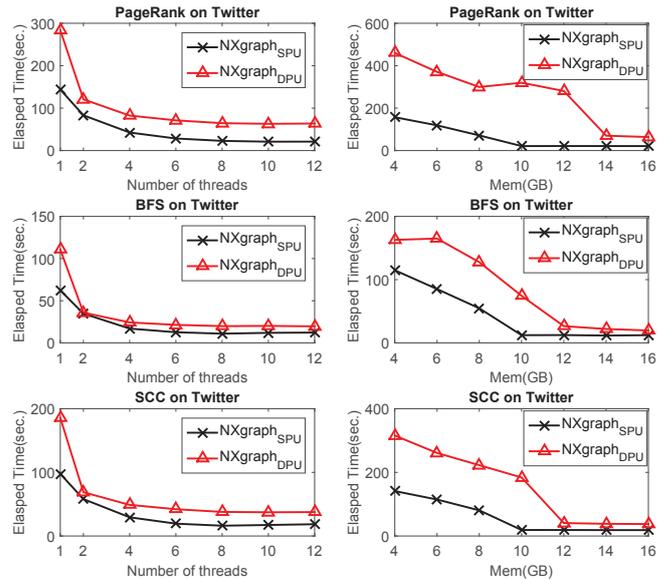

Fig. 8: SPU vs DPU on performance.

[21]. GraphChi [22] adopts another sorting policy, namely sort the edges by sources, so that its coarse-grained Parallel Sliding Windows (PSW) model can apply. GridGraph [30] chooses not to sort the edges at all. In both of the last two cases, parallelism has to be enabled by a coarse-grained model. We test both sub-shard ordering and parallelism models with the three real-world graphs, as listed in Table IV. As shown from the results, the destination-ordering and fine-grained parallelism always outperforms the other in all three assessed cases.

*2) Exp 2 Partitioning:* Figure 7 demonstrates the impact of partitioning on system performance for different algorithms. Global query algorithms like PageRank are less sensitive to the number of intervals and sub-shards, whereas targeted query algorithms like SCC are more sensitive because active status is saved in unit of intervals. Too less number of intervals will make it hard to efficiently skip unnecessary sub-shards. For global queries, an enough number of intervals are necessary to enable full overlap among execution threads and maintain high-parallelism. However, too many number of sub-shards will bring more overhead of thread switching and sub-shard storage. From the experimental results, $P = 12$ to $48$ are all good practices for interval partitioning.

*3) Exp 3 SPU vs DPU:* Figure 8 compares the performance of SPU and DPU under different environments and algorithms. As shown in the figure, SPU always outperforms DPU in all assessed cases, demonstrating the advantages of SPU scheme. Therefore, SPU is always preferred over DPU. As a combination of SPU and DPU, MPU will try to apply SPU on as many sub-shards as possible, as described in Section III-B3.

### C. Different Environments

In this subsection, 10 iterations of PageRank are performed on three real-world graphs respectively, namely Live-journal [32], Twitter [1] and Yahoo-web [31].

*1) Exp 4 Memory Size:* Experiment 1 reports the elapsed time as memory size changes. The number of threads is 12 in this experiment. As can be seen in Figure 9, for small graphs like Live-journal, performance hardly changes as all intervals and sub-shards can be stored in memory. In this case, NXgraph outperforms TurboGraph and GraphChi by fully utilizing CPU locality and enabling high parallelism. For Twitter graph, performance of NXgraph saturates at about 10 GB of memory, which is approximately the point when all intervals and sub-shards are loaded into memory. For larger graph, namely Yahoo-web, the saturation point is about 40GB.

*2) Exp 5 Number of Threads:* Experiment 2 reports the elapsed time as the number of available threads changes. The memory size is 16GB in this experiment. As can be seen in Figure 10, for relatively small graphs like Live-journal and Twitter whose data can be completely loaded in memory, the degree of parallelism has significant impact on system performance of NXgraph because NXgraph can exploit the degree of parallelism very well. However, in the case of Yahoo-web, performance of NXgraph is limited by disk I/O. Thread number has relatively less impact.

### D. Different Tasks

*1) Exp 6 Scalability:* Experiment 3 reports the flexible scalability of NXgraph. As analyzed in Section III-B, the relative amount of I/O does not increase as the scale of graph increases. In a certain range, larger graph can bring higher parallelism since the average overhead of thread switching can be reduced. This is shown in Figure 11. As the scale of graph changes in this experiment, execution time cannot represent system performance very well. Instead, Million Traversed Edges Per Second (MTEPS) is reported as a metric of system performance, indicating the system throughput. It can be observed that TurboGraph shows a tendency of decrease on throughput, which matches the analysis in Section III-C. GraphChi does not show a clear tendency of increase or decrease on performance in the assessed range of graph scale.

*2) Exp 7 More Tasks:* Experiment 4 reports the execution time of more graph algorithms on the three real-world graphs. These algorithms includes Breadth-First Search (BFS), Strongly Connected Components (SCC) and Weakly Connected Components (WCC). We use full computational resources in this experiment, namely, 12 threads and 64GB memory. Although performed on the basis of iterations, NXgraph can skip unnecessary updates on certain sub-shards by setting the active status of its corresponding intervals as described in Section II-B, which makes it as efficient even for algorithms like BFS and connected components. However, as can be seen in the third subfigure, TurboGraph outperforms NXgraph on WCC of the Yahoo-web. This is because TurboGraph is fully optimized for targeted queries by using very small intervals (several megabytes) as analyzed in their work [23]. For Yahoo-web, such small intervals will result in thousands of intervals and millions of sub-shards. NXgraph currently does not support such large number of sub-shards due to the limitation on the number of opened files by the operating system.

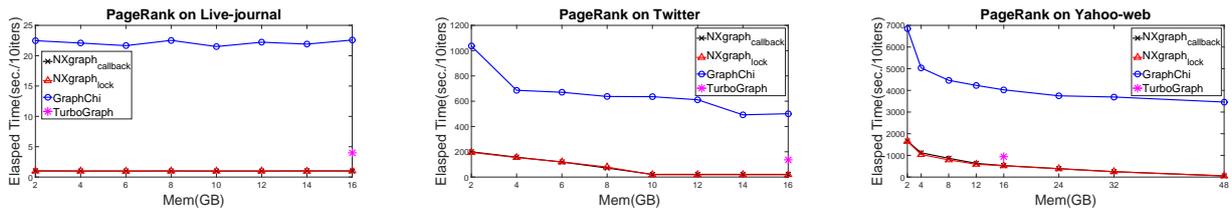

Fig. 9: Memory size changes (12 threads). We cannot change the memory size on Windows for TurboGraph.

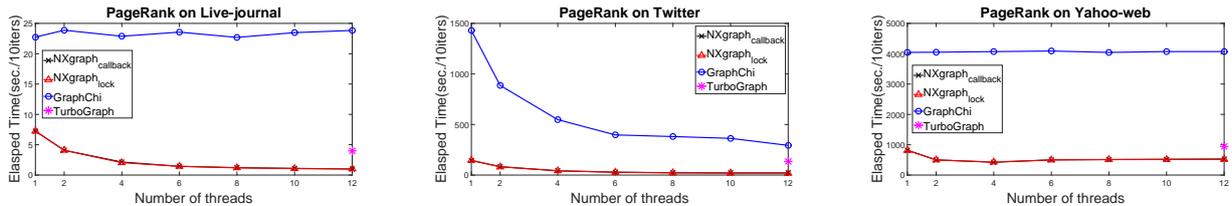

Fig. 10: Number of threads changes (16GB memory). We cannot change the number of threads on Windows for TurboGraph.

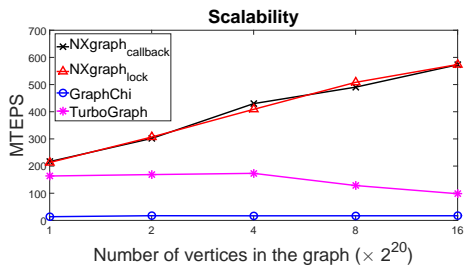

Fig. 11: Scalability.

Note that TurboGraph does not provide a program for SCC nor an interface to implement one, therefore we cannot assess its performance for that. Besides, BFS program provided by TurboGraph keeps crashing so that we are not able to access its performance, either. GraphChi cannot finish SCC or WCC on Yahoo-web within 24 hours. We set the root vertex to the first one for all graphs and systems in the BFS algorithm.

### E. Comparison to Other Systems

In Experiment 5 and 6, we run NXgraph under two different resources availabilities and compares it to other state-of-the-art graph processing systems that are not included in our own experiments. Performance reported in this section are cited from previous work except for NXgraph. The "evaluation environment" column in the table reports the CPU resources, number of available threads, memory budget size/physical memory size and the type of disk in the above order.

*1) Exp 8 Limited Resources:* In Table V, NXgraph is assessed under HDD with 8 available threads and 8GB memory, which simulates the evaluation setup of VENUS [21]. Since we cannot access the executable of VENUS or the source code, we compare NXgraph with VENUS in this way. As can been seen in the table, NXgraph (HDD) outperforms VENUS by about 7.6 times. The reason why NXgraph outperforms VENUS is analyzed as follows.

The v-shard is the key idea involved in VENUS. A v-shard is composed of all in-neighbors of vertices in an interval. It is very efficient to perform update with the presence of v-shards, which significantly reduce the amount of data transfer by reading vertex attributes from the in-memory v-shard. However, a v-shards is usually much larger than its corresponding interval. A vertex is very likely to have connection with any other vertex, in or out of its own interval. A v-shard much fit in memory to bring benefits and this results in even smaller intervals. With trivial effort on preprocessing, a large amount of intervals will be generated. This will then result in large overhead to maintain consistency of vertex attributes in each v-shard. To obtain high performance under VENUS framework, partitioning must be carefully designed so that connections between intervals are minimized, which is a non-trivial clustering problem. NXgraph does not introduce the v-shards to reduce the amount of data transfer or to maintain locality.

The other four results in Table V are reported by GridGraph in [30]. These results are run in an AWS EC2 server with a high volume of physical memory, which might be used for cache by the operating system. Note that VENUS states that the file caching mechanism is disabled in their experiments but we are not able to disable it successfully with the same tools.

*2) Exp 9 Best Performance:* In Table VI, NXgraph is assessed under 8 threads and 16GB memory on SSD. Further increase resources hardly improves performance for the assessed task. We also report the best case performance obtained

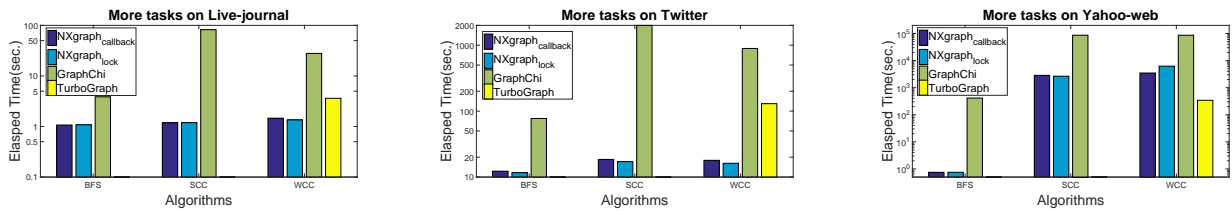

Fig. 12: BFS, SCC, and WCC on Live-journal, Twitter, and Yahoo-web.

TABLE V: System performance with limited resources

| System | Time(s) | Speedup | Evaluation environment |
|---|---|---|---|
| NXgraph | 7.13 | 1.00 | Intel i7 3.3GHz, 8t, 8G, SSD |
| GridGraph | 26.91 [30] | 3.77 | AWS EC2 8t, 8G/30.5G, SSD |
| X-stream | 88.95 [30] | 12.48 | AWS EC2, 8t, 8G/30.5G, SSD |
| NXgraph | 12.55 | 1.00 | Intel i7 3.3GHz, 8t, 8G, HDD |
| VENUS | 95.48 [21] | 7.60 | Intel i7 3.4GHz, 8t, 8G/16G, HDD |
| GridGraph | 24.11 [30] | 1.92 | AWS EC2, 8t, 8G/30.5G, HDD |
| X-stream | 81.70 [30] | 6.51 | AWS EC2, 8t, 8G/30.5G, HDD |

Task: 1 iteration of PageRank on Twitter [1] graph

TABLE VI: System performance in the best case

| System | Time(s) | Speedup | Evaluation environment |
|---|---|---|---|
| NXgraph | 2.05 | 1.00 | Intel i7 3.3GHz, 8t, 16G, SSD |
| X-stream | 23.25 [28] | 11.57 | AMD Opteron 2.1GHz, 32t, 64G, SSD |
| GridGraph | 24.11 [30] | 11.99 | AWS EC2, 8t, 8G/30.5G, HDD |
| MMAP | 13.10 [33] | 6.52 | Intel i7 3.5GHz, 8t, 16G/32G, SSD |
| PowerGraph | 3.60 [18] | 1.79 | 64×(AWS EC2 Intel Xeon, 16t, 23G) |

Task: 1 iteration of PageRank on Twitter [1] graph

by other graph processing systems on the same task. As shown in the table, NXgraph outperforms all assessed single-machine system even with better resources. Moreover, NXgraph is able to outperform PowerGraph by 1.79 times, which is a distributed system consists of a cluster of 64 high-performance machines.

## V. RELATED WORK

Previous graph processing work can be classified into two categories, distributed systems and single-machine systems. Distributed systems use a cluster of machines for large-scale graph problems. These systems need to handle synchronization and consistency problems among the machines. In contrast, single-machine systems can solve large-scale graph problems in a comparable time against distributed systems, with lower power consumption and lighter communication cost.

### A. Distributed Systems

Pregel [16] is a distributed synchronous system for large graphs. Pregel applies the vertex-centric programming model. In this model, each vertex executes a kernel function to update its neighbor vertices. Meanwhile, Pregel adopts the Bulk-Synchronous Parallel (BSP) model. In this model, kernel function is executed in unit of super-steps. Kernels functions of different vertices are executed in parallel within a super-step. A barrier is imposed between super-steps so that the whole execution is synchronized.

GraphLab [17] is an asynchronous distributed system designed for machine learning algorithms. GraphLab applies the vertex-centric model. Each vertex executes the updating algorithm and is able to access graph data on other machines. PowerGraph [18] is another asynchronous distributed system, focusing on the partitioning of large-scale graph in a distributed graph processing systems. In the PowerGraph system, each vertex is attached to its master machine and its mirrors will be maintained on all other machines. All mirrors will be sent to the master machine to update the vertices, which brings communication overhead.

Synchronous systems suffer from the imbalanced load among different machines while asynchronous systems make lots of effort to guarantee the data consistency of different machines. All these distributed graph processing systems suffer from the overhead of fault tolerance, low robustness and heavy communication costs. In this paper, we compare the distributed systems with NXgraph on certain computational tasks. The result shows that NXgraph is 1.79x faster than PowerGraph when performing PageRank on the Twitter [1] graph.

### B. Single-machine Systems

GraphChi [22] is the first graph processing system based on the interval-shard structure on a single machine. It applies the Parallel Sliding Windows (PSW) method for processing very large graphs from disk. Vertices are divided into intervals and exactly one shard is attached to each interval. The shard consists of all edges with destination vertices in its associated interval. Edges are sorted by their source vertex indices so that PSW model can apply. The interval-shard structure ensures data access locality. The sliding windows ensures streamlined data access for each shard but among different shards, the access pattern is random. Besides, all related edges need to be loaded into memory in GraphChi. Therefore, GraphChi requires relatively more disk data transfer. Another difference between GraphChi and NXgraph is that GraphChi does not provide high degree of parallelism, limited by its high disk data transfer and randomness of I/O.

TurboGraph is another disk-based graph processing system, which proposes a novel parallel model, *pin-and-slide*. It contains a list of slotted pages. Each page contains the outgoing edges of several vertices. TurboGraph system uses a buffer pool in the memory to store several pages. It also divides the vertices into several *chunks* to ensure data access locality. Each chunk is loaded into memory in sequence and updated by each edge in the buffer pool. TurboGraph requires SSD to ensure its performance because it uses parallel I/O. Compared with TurboGraph, NXgraph reduces the amount of data transfer and enables streamlined disk access pattern.

VENUS [21] is more friendly to hard disks. It enables Vertex-centric Streamlined Processing (VSP) on their system. The system proposes *v-shards* to store the source vertices of edges in each shard. VENUS provides two algorithms with different implementations of v-shard. The first algorithm, called VSP-I, materializes all v-shard values in each shard. And the second algorithm, called VSP-II, applies *merge-join* to construct all v-shard value on-the-fly. By involving the v-shards, VENUS enables streamlined disk I/O with fine-grained parallelism. NXgraph achieves streamlined disk access pattern without introducing v-shards. Another difference between VENUS and NXgraph is that sub-shards in NXgraph achieve a higher locality than v-shards in VENUS.

GridGraph [30] adopts similar streamlined processing model on a single machine. In the GridGraph system, edges are further divided into smaller grids rather than shards in the GraphChi system. Meanwhile, GridGraph applies a 2-level hierarchical partitioning of the grids, which organizes several adjacent grids into a larger virtual grid. In this way, GridGraph can not only ensure data locality but also reduce the amount of disk I/O. However, with a TurboGraph-like updating strategy, GridGraph requires more disk data transfer than NXgraph. Besides, GridGraph can not fully utilize the parallelism of multi-thread CPU without sorted edges.

## VI. CONCLUSION

We have presented NXgraph, an efficient graph computation system that is able to handle web-scale graphs on just a single machine. It provides three novel update strategies under the Destination-Sorted Sub-Shard (DSSS) structure. It also has an efficient implementation fully utilizing the main memory, CPU cache locality and parallelism, reducing the amount of disk I/O. Extensive experiments on three real-world graphs and five synthetic graphs show that NXgraph can outperform GraphChi, TurboGraph, VENUS and GridGraph in various situations. Moreover, NXgraph, running on a single commodity PC, can outperform PowerGraph, a distributed graph processing system for PageRank on the Twitter graph. For future work, NXgraph will be extended to support dynamic change on graph structure, which will make NXgraph capable of more graph computation tasks. We will also try to optimize our system for some representative graph algorithms, e.g., more intervals for targeted queries may lead to even better performance.